\documentclass[sigconf]{acmart}

\usepackage{amsmath}
\usepackage{amsfonts}
\usepackage{amssymb}
\usepackage{calligra}
\usepackage{pifont}

\usepackage{graphicx}
\usepackage{adjustbox}
\usepackage{listings}
\usepackage{multirow}
\usepackage{color}
\usepackage{tcolorbox}
\usepackage{footnote}
\usepackage{cleveref}
\usepackage{tabularx, tabulary}
\usepackage{todonotes}

\usepackage[noend]{algpseudocode}
\usepackage{algorithm}
\algrenewcommand\algorithmiccomment[2][\itshape]{{#1\hfill\(\triangleright\)
    #2}}
\algrenewcommand{\algorithmicrequire}{\textbf{Input:}}
\algrenewcommand{\algorithmicensure}{\textbf{Output:}}

\usepackage[labelformat=simple]{subcaption}

\usepackage{tikz}
\usetikzlibrary{arrows,decorations.text,backgrounds,shadows,calc}
\usetikzlibrary{intersections, patterns,automata,arrows,positioning}
\usetikzlibrary{matrix}
\usepackage{varwidth}
\usetikzlibrary{external}
\usepackage{standalone}
\usepackage{tablefootnote}

\tikzstyle{ca}=[draw,circle,fill=black!5,rounded corners]  
\tikzstyle{caf}=[draw,double, circle,fill=black!5,rounded corners] 
\tikzstyle{leer} = [rectangle,node distance=.6cm]              

\tikzset{>=latex}
\tikzstyle{ent}=[draw, fill=blue!20, text width=6.8em,
    text centered,minimum width=4em, minimum height=3em, rounded corners, drop shadow]
\tikzstyle{inv}=[draw, fill=blue!20, text width=4em,
text centered,minimum width=4em, minimum height=1em, rounded corners, drop shadow]
\tikzstyle{ent-small}=[draw, fill=blue!20, text width=3em,
    text centered,minimum width=3em, minimum height=2em, rounded corners, drop shadow]
\tikzstyle{ent-large}=[draw, fill=blue!40, text width=8em,
    text centered,minimum width=8em, minimum height=4em, rounded corners, drop shadow,
     rectangle split, rectangle split draw splits=false, rectangle split parts=2]

 \newcommand{\llbox}[1]{
	\begin{tcolorbox}[width=\columnwidth, colframe=black, boxrule=0.25mm, top=1mm, left=1mm, right=1mm, bottom=1mm]
		#1
	\end{tcolorbox}
}

 \usepackage{pgfplots}
 \pgfplotsset{
 	compat=1.9,
 	log ticks with fixed point, 
 	table/col sep=tab, 
 	unbounded coords=jump, 
 	filter discard warning=false, 
 }
    
\definecolor{keyword-color}{RGB}{127,0,85}
\definecolor{dark-green}{RGB}{0,112,0}

\newcommand{\NoteEnv}[3]{\newenvironment{#1}{\color{#3}#2: }{}}
\definecolor{jancolor}{rgb}{0.9,0,0.0}
\definecolor{heikecolor}{rgb}{0.8,0.3,0}
\definecolor{var}{rgb}{0.4,0.1,0.1}

\NoteEnv{jan}{Jan}{jancolor}
\NoteEnv{heike}{Heike}{heikecolor}

\newcommand{\IG}{helper invariant generator}
\newcommand{\InvGen}{Helper Invariant Generator}
\newcommand{\master}{master verifier}
\newcommand{\Master}{Master Verifier}

\newcommand{\CFA}{$C$}

\newcommand{\safetyProp}{$P$}

\newcommand{\witness}{$\omega$}
\newcommand{\witnessPrime}{$\omega'$}
\newcommand{\cpachecker}{\textsc{CPAchecker}}
\newcommand{\ua}{\textsc{UltimateAutomizer}}
\newcommand{\veriabs}{\textsc{VeriAbs}}
\newcommand{\seahorn}{\textsc{SeaHorn}}
\newcommand{\covercig}{CoVerCIG}

\newcommand{\svcomp}{SV-COMP}
\newcommand{\android}{Android}

\newcommand{\Var}{\mathit{Var}}

\newcommand{\benchexec}{\textsc{Benchexec}}

\newsavebox\mybox

\AtBeginDocument{%
	\providecommand\BibTeX{{%
			\normalfont B\kern-0.5em{\scshape i\kern-0.25em b}\kern-0.8em\TeX}}}

\copyrightyear{2020}
\acmDOI{}
\begin{document}
	
\title{Cooperative Verification via Collective Invariant Generation}

\author{Jan Haltermann}

\authornote{This author was partially supported by the German Research Foundation (DFG)
	under contract 418257054.}
\affiliation{%
	\institution{Paderborn University}
	\streetaddress{Warburger Str. 100}
	\city{Paderborn}
	\country{Germany}
	\postcode{33098}
}\email{jfh@mail.upb.de}
\author{Heike Wehrheim}

 \affiliation{%
 	\institution{Paderborn University}
 	\streetaddress{Warburger Str. 100}
 	\city{Paderborn}
 	\country{Germany}
 	\postcode{33098}
}
\email{wehrheim@upb.de}
	
%
%

\renewcommand{\shortauthors}{Haltermann and Wehrheim}

\begin{abstract}
  Software verification has recently made enormous progress due to the development of novel verification methods and the speed-up of 
  supporting technologies like SMT solving. To keep software verification tools up to date with these advances, tool developers 
  keep on integrating newly designed methods into their tools, almost exclusively by re-implementing the method within their own framework. 
  While this allows for a conceptual re-use of methods, it requires novel implementations for every new  technique. 

  In this paper, we employ {\em cooperative verification} in order to avoid re-implementation and enable usage of 
  novel tools as black-box components in verification. Specifically, cooperation is employed for the core ingredient of 
  software verification which is {\em invariant generation}. Finding an adequate loop invariant is key to the success of 
  a verification run. Our framework named \covercig \ allows a master verification tool to delegate the task of invariant generation to 
  one or several specialized helper invariant generators. Their results are then utilized within the verification run of the 
  master verifier, allowing in particular for crosschecking the validity of the invariant. 
  We experimentally evaluate our framework on an instance with two masters and three different invariant generators 
  using a number of benchmarks from \svcomp \ 2020.
  The experiments show that the use of \covercig \ can increase the number of correctly verified tasks without increasing the used resources.

\end{abstract}
  	\keywords{Cooperation, Software Verification, Invariant Generation}	

\ccsdesc[500]{Software and its engineering~Formal software verification}
	

	\maketitle

	
\section{Introduction} 

Recent years have seen a major progress in software verification as for instance witnessed by the annual competition on software 
verification SV-COMP~\cite{DBLP:conf/tacas/Beyer17}. This success is on the one hand due to advances in SAT and SMT 
solving and 
on the other hand due to novel verification methods like interpolation in model checking~\cite{DBLP:reference/mc/McMillan18}, automata-based 
software verification~\cite{DBLP:conf/cav/HeizmannHP13} or property directed reachability~\cite{DBLP:conf/vmcai/Bradley11}.  
Still, automatic verification remains a complex and error-prone task. 
In particular, it is often the case that one tool can verify a particular class of programs, but fails to verify other classes (or even 
gives incorrect answers), whereas it is the reverse situation for another tool. 
Moreover, to keep their tools up to date with novel techniques, tool developers 
keep on integrating them by re-implementation within their framework. 

An approach for changing this unsatisfactory situation is 
{\em cooperative verification} (for an overview see~\cite{DBLP:journals/corr/abs-1905-08505}). Cooperative verification 
builds on the idea of letting tools (and thus techniques) cooperate on verification tasks, thereby leveraging the tool's 
individual strengths. In particular, cooperative verification aims at {\em black box} combinations 
of tools, using existing tools off-the-shelf without re-implementation. 
While this sounds like a natural idea, its realization poses a number of challenges, the major one 
being the {\em exchange} and {\em usage} of analysis information. For cooperation, tools are required to produce 
(partial) results which other tools can understand and employ in their verification run. 
With conditional model checking~\cite{DBLP:conf/sigsoft/BeyerHKW12}, the first proposal of an exchange
 {\em format} for  verification results was made. A conditional model checker outputs its (potentially partial) 
result in the form of a {\em condition} which can be read by other conditional model checkers 
in order to complete the verification task. Since verification tools normally do not understand conditions, 
{\em reducers}~\cite{Czech2015,DBLP:conf/icse/BeyerJLW18} have been proposed to bring conditions 
back into a form understandable by verifiers, namely into (residual) programs describing the so far unverified 
program part.  
This allows the result of a conditional model checker to be made usable by arbitrary other verifiers. 
A second type of existing result usage is the {\em validation} of tool's results~\cite{DBLP:conf/sigsoft/0001DDHS15,DBLP:conf/spin/JakobsW14}, 
similar to proof-carrying code~\cite{DBLP:conf/popl/Necula97}. 
Both of these types are {\em sequential} forms of cooperation: a first verifier starts and a second verifier continues, 
either by completing or by validating a first result. 

\begin{figure}[th]
  
  \adjustbox{max width=\linewidth}{
    \begin{tikzpicture}[node distance=7em]
      \node[ent-large, name=veri](veri) {\begin{tabular}{c} {\bf \large Master} \\ {\bf \large Verifier} \end{tabular}\nodepart{two}WitnessInjector};
      \draw[dashed] (veri.text split west) -- (veri.text split east);
      \node[ent-small,below left of=veri,xshift=-10pt] (ada1) {Adapter};
      \node[ent,below left of=ada1] (inv1) {\begin{tabular}{c}  Helper Invari- \\ ant Generator \end{tabular}};
      \node[ent-small,below right of=veri,xshift=10pt] (ada2) {Adapter};
      \node[ent,below right of=ada2] (inv2) {\begin{tabular}{c}  Helper Invari- \\ ant Generator \end{tabular}};
      \node[ent-small,left of=ada1] (map1) {Mapper};
      \node[ent-small,right of=ada2] (map2) {Mapper};
      \draw[->] ($(inv1.north west)!0.75!(inv1.north east)$) -- node[right,pos=.4] {\ Invariant} ($(ada1.south west)!0.2!(ada1.south east)$);  
      \draw[->]   (ada1) -- node[right,pos=.4] {Witness}  (veri);  
      \draw[->] ($(inv2.north west)!0.25!(inv2.north east)$) -- node[left,pos=.4] {Invariant\ } ($(ada2.south west)!0.8!(ada2.south east)$) ;  
      \draw[->]   (ada2) -- node[left,pos=.4] {Witness}  (veri);  
      \draw[->] (veri) -- node[above, sloped] {Prog+Prop} (map1);
      \draw[->] (veri) -- node[above, sloped] {Prog+Prop} (map2);
      \draw[->] (map1) -- node[left] {Task} (inv1);
      \draw[->] (map2) -- node[right] {Task} (inv2);
      \path (veri.south) +(0,-2.6) node  (cv) {\Huge{$|| \quad \ldots \quad ||$}};
      \node[left of=veri,xshift=-4.1cm,yshift=.6cm] (f1) {};
      \node[left of=veri,xshift=-4.1cm,yshift=-.6cm] (f2) {};
      \path [draw,->] (f1) -- node[above,pos=.12,sloped] {Program} (veri);
      \path [draw,->] (f2) -- node[above,pos=.12,sloped] {Property} (veri);
      \node[right of=veri,xshift=3.9cm] (f2) {};
      \path [draw,->] (veri.east) -- node[above,pos=.86]{Result} (f2); 
      \begin{pgfonlayer}{background}
        \path (veri.west |- veri.north)+(-3.5,0.5) node (a) {};
        \path (veri.south -| veri.east)+(+3.5,-3.5) node (c) {};
        \path[fill=yellow!20,rounded corners, draw=black!50]
            (a) rectangle (c);
      \end{pgfonlayer}
    \end{tikzpicture}
  } 
  \caption{Collective invariant generation}
  \label{fig:approach}

\end{figure}
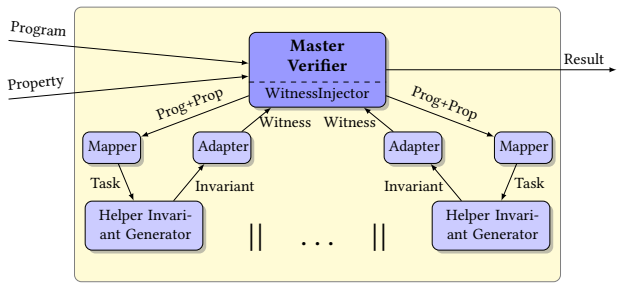

\smallskip 
\noindent In this paper, we propose \covercig, a cooperation framework which complements these existing approaches 
by a new type of cooperation. Conceptually, this framework (depicted in Figure~\ref{fig:approach}) 
consists of a {\em master verifier} and a number of {\em \IG s}. 
The master verifier has the overall control on the verification process and can {\em delegate} tasks to helpers as well as {\em continue} 
its own verification process with (partial) results provided by helpers. 
The helpers run in parallel as black boxes without cooperation. 
The task to be delegated is an integral part of software verification, namely {\em invariant generation}. 
The framework  allows cooperation via outsourcing the task of invariant generation, 
leveraging the strength of specialized invariant generation tools.  

Like for other types of cooperation, the question of the exchange format for results comes up. 
Here, we have chosen {\em correctness witnesses}~\cite{DBLP:conf/sigsoft/0001DDH16} for this purpose. 
Correctness witnesses are employed in witness validation and certify a verifier's result stating the correctness 
of a program. These witnesses are particularly well suited for our intended usage, because their format 
is standardized and a number of verifiers already produce correctness witnesses. 
To account  for the incooperation of helper verifiers not producing witnesses, our framework also 
foresees the 
inclusion of {\em adapters} transforming invariants into correctness witnesses. We provide an implementation of two 
such adapters. Witnesses are then {\em injected} into the verification run of the master. 
For stating the task to be solved by invariant generators 
we furthermore require {\em mappers} transforming program and property to be proven into a task format 
understandable by the helper tools. 
Figure~\ref{fig:approach} depicts our framework for collective invariant generation. 
The framework can be arbitrarily configured with different masters and helpers, provided that suitable adapters and mappers 
are given. 

We have implemented our framework within the \cpachecker \ framework~\cite{DBLP:conf/cav/BeyerK11}
and have employed different configurations of it as master verifier. As helper verifiers we have 
chosen publicly available verification tools, some producing and one not producing witnesses. 
We have then experimentally evaluated 14  different combinations of master and helper on benchmarks of the annual competition 
of software verification \svcomp~\cite{DBLP:conf/tacas/Beyer17}. 
The experiments show an improvement over the verification capabilities of the master tool, 
without incurring significant overhead. In some cases, the verification time is even decreased in 
cooperative verification. 

\smallskip
\noindent
Summarizing, we make the following contributions.
\begin{itemize}
  \item We propose a framework for cooperative software verification based on a master-helper architecture for 
    collective invariant generation. 
  \item We construct 14 different instantiations of the framework using 2 masters and 3 helpers, running both 
    helpers in isolation as well as in parallel. 
  \item For the inclusion of  helper verifiers, we implement two adapters, one transforming invariants 
    expressed in the LLVM IR language\footnote{https://llvm.org/docs/LangRef.html} into correctness witnesses, the other 
   modifying a generated witness as to bring it into the right format. 
  \item We carry out an extensive experimental evaluation demonstrating the effectiveness and efficiency of collective 
    invariant generation. 
\end{itemize}

\section{Fundamentals}
We aim at the cooperative verification of programs written in GNU C, focusing on the validation of safety properties. 
To be able to define safety properties, a formal representation of programs as well as their semantics is needed. 
Thus we next briefly introduce the syntax and semantics of programs which we consider here.

We follow the notation of Beyer et al.~\cite{DBLP:reference/mc/BeyerGS18} describing programs as 
{\em control-flow automata} (CFAs). 
A CFA is basically a control-flow graph with edges annotated with program statements.  
More formally, a program is represented as a control-flow automaton \linebreak$C= (L,l_0,G)$, consisting of a set of program locations $L$, an initial location $l_0 \in L$ and the control-flow edges $G, G \subseteq L \times Op \times L$.
The set $Op$  contains all possible operations on integer variables\footnote{In our formalization, we use integer variables only, the 
implementation covers C programs.} present in the program, namely  conditions (as of conditionals and loops), assignments, method 
calls and return statements.
\Cref{fig:example-a} shows a C-program taken from the \svcomp \ benchmarks\footnote{https://github.com/sosy-lab/sv-benchmarks}, and \Cref{fig:example-b} its corresponding CFA. The program also contains a special {\em error} label, used for encoding the property 
to be verified. The verification task for this program is to show the non-reachability of the error label at location 9, i.e., for our example program the verifier has to 
prove that $y$ equals $n$ after the loop which is true (since $n$ is unsigned). 

For the semantics, we start by defining program states. 
Let $\Var$ denote the set of all integer variables occurring in programs, $BExp$ the set of boolean expressions 
and $AExp$ the set of arithmetic expressions over $\Var$.
Then a \textit{state} $\sigma$ of the program  is a mapping from  the variables to the integers, i.e., $\sigma: \Var \rightarrow \mathbb{Z}$. 
We lift the mapping to also contain the evaluation of arithmetic and boolean expressions so that $\sigma$ maps 
$AExp$ to $\mathbb{Z}$ and $BExp$ to $\mathbb{B}$.
A finite \textit{program path $\pi$} is a sequence of \textit{transitions} $\langle \sigma_0, l_0 \rangle \overset{g_0}{\rightarrow} \langle \sigma_1, l_1 \rangle \cdots \overset{g_{n-1}}{\rightarrow} \langle \sigma_n, l_n \rangle$, such that $\sigma_0$ assigns 0 to all variables, $l_n$ is a leaf in the CFA and 
$(l_i, g_i, l_{i+1}) \in G$ holds for each transition $\langle \sigma_i, l_i \rangle \overset{g_i}{\rightarrow} \langle \sigma_{i+1}, l_{i+1} \rangle $ in $\pi$.
Infinite program paths are defined analogeously.
As for state changes in paths: If $g_i$ is a boolean expression, method call or return statement, then $\sigma_i = \sigma_{i+1}$ holds. 
If $g_i$ is an assignment $x=a$, where $a \in AExp$, then $\sigma_{i+1} = \sigma_i[x \mapsto \sigma_i(a)]$. 
Finally, we denote all paths of a program represented by a CFA \CFA \ by $paths($\CFA$)$.\\

\savebox{\mybox}{	\begin{tikzpicture}[>=stealth',shorten >=1pt,auto, node distance=1.5 cm, scale = .9, transform shape]
	\node[state,initial] 	(l1)   {$q_1$}; 
	\node[state] 			(l2) 	[below of=l1]  {$q_2$};
	\node[state]			(l4) 	[below of=l2]  {$q_3$};
	\node[state]			(l6) 	[below of=l4]  {$q_4$};
	\node[state]			(l7) 	[right=1.5cm of l4]  {$q_5$};
	\node[state]			(l8) 	[right of=l6]  {$q_6$};
	\node[state]			(l9) 	[right  of =l8]  {$q_7$};
	\node[fill=black!15!green, text centered, text depth = .05\baselineskip] (w4) 	[right of=l2 ] {n == x+y };
	
	\path[->] 
	(l1) edge  node {1,enterFunc} (l2)
	(l2) edge [left]  node {3,enterLoopHead} (l4)
	(l4) edge [above] node {4,else} (l7)
	(l4) edge [right, bend left=45,pos=.6]  node {4,then} (l6)
	(l6) edge[left,pos=.3] node {6,enterLoopHead} (l4)
	(l7) edge [left,pos=.15] node {7,then} (l8)
	(l7) edge  node {7,else} (l9)
	(l1) edge [loop right] node {o/w} (l1)
	(l2) edge [loop left] node {o/w} (l2)
	(l4) edge [loop left] node {o/w} (l4)
	(l6) edge [loop below, right] node {o/w} (l6)
	(l7) edge [loop right] node {o/w} (l7)
	(l8) edge [loop below,right] node {o/w} (l8)
	(l9) edge [loop below,right] node {o/w} (l9)
	;
	\path[->]
	(w4) edge [dotted] (l4);
\end{tikzpicture}	
}
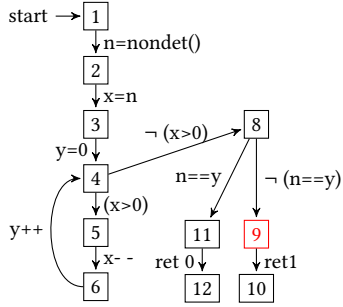
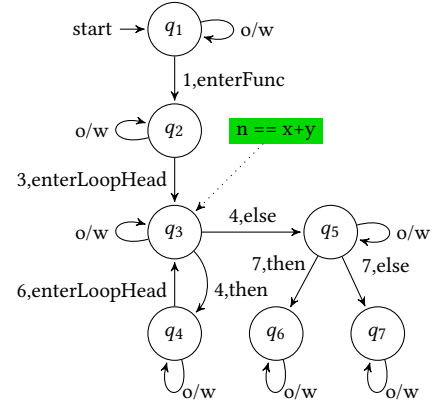
\begin{figure*}[th]
\begin{subfigure}{.33\textwidth}
  \vbox to \ht\mybox{%
	\vfill
	\begin{lstlisting}[tabsize=2, columns=flexible,mathescape=true,numbers=left,basicstyle=\ttfamily,	stepnumber=1, language=C]
int main() {
	unsigned int n = nondet();
	unsigned int x = n, y = 0;
	while(x > 0){	
		 x--;
		 y++; } 
  // Safety property
	if (!(n == y)) { 
		<@\textcolor{red}{Error: return 1;}@> }
	return 0;
}
\end{lstlisting}
\vfill}
\caption{C code example}
		\label{fig:example-a}
\end{subfigure}
\begin{subfigure}{.3\textwidth}
	  \vbox to \ht\mybox{%
		\vfill
		\begin{tikzpicture}[>=stealth',shorten >=1pt,auto,node distance=.8 cm, scale = .9, transform shape]
	\node[rectangle,draw,initial] 	(l1)   {1}; 
	\node[rectangle,draw] 			(l2) 	[below of=l1]  {2};
	\node[rectangle,draw] 			(l3) 	[below of=l2]  {3};
	\node[rectangle,draw] 			(l4) 	[below of=l3]  {4};
	\node[rectangle,draw] 			(l5) 	[below of=l4]  {5};
	\node[rectangle,draw] 			(l6) 	[below of=l5]  {6};
	\node[rectangle,draw] 			(l7) 	[right=2cm of l3]  {8};
	\node[rectangle,draw, red] 		(l9) 	[below=1.2cm  of l7]  {9};
	\node[rectangle,draw] 			(l8) 	[left of=l9]  {11};
	\node[rectangle,draw] 			(l10) 	[below of= l8]  {12};
	\node[rectangle,draw] 			(l11) 	[below of= l9]  {10};
	\path[->] 
	(l1) edge  node {n=nondet()} (l2)
	(l2) edge  node {x=n} (l3)
	(l3) edge [left]  node {y=0 } (l4)
	(l4) edge [above] node {$\neg$ (x>0) } (l7)
	(l4) edge  node { (x>0)} (l5)
	(l5) edge  node { x- -} (l6)
	(l6) edge[bend left=90] node {y++} (l4)
	(l7) edge [left] node {n==y} (l8)
	(l7) edge  node {$\neg$ (n==y)} (l9)
	(l8) edge  [left] node {ret 0} (l10)
	(l9) edge [] node {ret1 } (l11)
	;
	\end{tikzpicture}	
		\vfill
	}
	\caption{The corresponding CFA}	\label{fig:example-b}
\end{subfigure}
\begin{subfigure}{.3\textwidth}
 \usebox{\mybox}	
	\caption{Part of the witness}
	\label{fig:example-c}
\end{subfigure}
\caption{An example program, its control flow automaton and one witness}
\label{fig:example}		
\end{figure*}

Here, we are interested in verifying safety properties of programs given as CFAs. 
For the purpose of this paper, we define a {\em safety property} \safetyProp \ as a pair of a location $\ell \in L$ and a boolean condition $\varphi \in BExp$. 
There can be multiple safety properties required to hold in a program. 
For our example program of Figure~\ref{fig:example} the property is $(8,n=y)$.  
For the verifier this is encoded in the form 
\begin{verbatim}
  8:  if (!(n==y))
  9:      Error: return 1;
\end{verbatim}
Later, we will see that different verifers require different encodings of the property to be checked, 
and hence mappers need to be applied to translate property encodings.

A CFA (or program) \CFA \ \textit{violates a safety property} $P=(\ell,\varphi)$ when the program reaches location $\ell$ in a 
state which does not satisfy $\varphi$. 
More formally, \safetyProp \ is violated by \CFA, if there is some path $\pi \in paths($\CFA$)$,  $\pi = \langle \sigma_0, l_0 \rangle \overset{g_0}{\rightarrow} \langle \sigma_1, l_1 \rangle \cdots \overset{g_{n-1}}{\rightarrow} \langle \sigma_n, l_n \rangle$ 
and some $i$, $0 \leq i \leq n$, such that $\ell_i = \ell$ and $\sigma_i(\varphi)=\mathit{false}$. 

Cooperatively verifying safety of programs is achieved in our framework via collective (loop) invariant generation.
Syntactically, a \textit{loop invariant} is a boolean expression associated to a loop head.
A loop invariant needs to hold (1) before the first loop execution and (2) after each loop execution.
The expression $n = x+y$, for instance, is a loop invariant for the program in Figure~\ref{fig:example-a}, 
associated to the loop head at location $4$.
This loop invariant facilitates verification of the safety property, because in conjunction with the negated loop condition
and information about initial variable values it ensures $n$ to be equal to $y$ after the loop. 
Other valid loop invariants would be $x\geq 0$, $ n = 3 \Rightarrow  y \leq 5$ or $\mathit{true}$, 
which however all do not help in proving the safety property. 
Especially the loop invariant $\mathit{true}$ does not provide any information, as it always is a valid loop invariant.
Thus, we call it a \textit{trivial invariant}.

\begin{figure}[tb]
	\begin{lstlisting}[frame=single, tabsize=4, columns=flexible,mathescape=true]
<node id="q3">
	<data key="invariant">n == x+y</data>
	<data key="invariant.scope">main</data>
</node>
<edge source="q2" target="q3">
	<data key="enterLoopHead">true</data>
	<data key="startline">3</data>
	<data key="endline">3</data>
	...
</edge>
	\end{lstlisting}
	\caption{Excerpt of a correctness witness for the example	}
	\label{fig:example-witness}

\end{figure}

As stated before, we chose {\em witnesses} (more specifically, correctness witnesses) as exchange format during collective invariant generation.
Formally, a witness is a finite state automaton in which transitions are labelled with so called {\em source code guards} and states can be equipped with 
boolean expressions. When all these boolean expressions are either $\mathit{true}$ or $\mathit{false}$, we call the witness {\em trivial}. 
Source code guards are of the form \texttt{location,type} where \texttt{type} can be \texttt{then}, \texttt{else}, 
\texttt{enterFunc} and \texttt{enterLoopHead}. 
The guard \texttt{o/w} (otherwise) is used if a source code line does not match the other guards present.
Via these labels we can match transitions of the automaton with edges in the CFA. 

In~\Cref{fig:example-c}, we see a correctness witness for our example program. 
State $q_3$ is reached by transitions labelled \texttt{3,enterLoopHead} or \texttt{6,enterLoopHead} and thus corresponds to the 
loop head at program location 4. Associated with this state is the invariant  $n = x+y$.

Syntactically,  correctness witnesses are stored in an XML format and consist of two parts: (1) general information like the producer of the witness or the program associated with the witness, and (2) a GraphML representation of the witness automaton. 
Figure~\ref{fig:example-witness}   shows an excerpt of this format for the witness in Figure~\ref{fig:example-c}. 
More information and a formal specification of correctness witnesses can be found in~\cite{DBLP:conf/sigsoft/0001DDH16}.

\section{Concept}  
In this section, we introduce our novel concept of \textbf{Co}operative \textbf{Ver}ification via \textbf{C}ollective \textbf{I}nvariant \textbf{G}eneration (\covercig), shown in \Cref{fig:approach}.
The framework contains two sorts of main components: Master verifiers (one) and helper invariant generators (several). 
Next, we state some requirements on and explain the functionality of these components as well as their cooperation.
\subsection{Components of the \covercig-Framework}
The most important component of the framework is the \master, which we build out of an existing verifier.   
The master is responsible for coordinating the verification process and can, if needed, request support from the second type of components, 
the helpers,  in the form of invariants as described by correctness witnesses.
Hence, the master is also steering the cooperation. 

In the following, we explain the two sorts of main components in more detail:
\begin{description}
	\item[\Master] A \textit{\master} gets as input the program \CFA \ as CFA and a safety property \safetyProp.
	It computes as output a boolean answer $b$, stating whether the property holds, and possibly (but not necessarily) provides 
        an overall witness \witness.
	To be able to process the provided support in form of invariants stored inside of correctness witnesses, a master is required to implement an internal function called \textit{injectWitness}.
	The function loads a witness, extracts the invariants present in it and injects them into the analysis of the \master.
	The witness injection can either happen before \mbox{(re-)starting} the analysis or during runtime.
	We exemplify the realization of witness injection later.
	\item[\InvGen] A \textit{\IG} \linebreak  gets as input the program \CFA \ as CFA and a safety property \safetyProp.
	It computes as output a set of invariants, stored in a verification witness \witnessPrime.
	The generated invariants are neither required to be helpful for the \master \ nor to be correct. 
	Thus, \IG s are also allowed to generate trivial invariants or invariant candidates which might turn out to be wrong. 
\end{description}


\begin{figure}[t]

\adjustbox{max width=\linewidth}{
		\begin{tikzpicture}
		\node[ent,  minimum height = 5em, text width= 7.2em](IR) {Invariants over IR-variables with IR-locations};
		\node[ent,  minimum height = 5em, text width= 7.2em ](C) [right=5em of IR]{Invariants over C-variables with C-locations};
		\node[ent,  minimum height = 5em, text width= 4em](witness) [right=5em of C] {witness};
		
		\draw[->] 
		(IR)  edge [above] node{translate} (C)
		(C) edge [above] node {construct} (witness);
		\end{tikzpicture}
	}
	\caption{Workflow of an adapter}
	\label{fig:adapter}
	\vspace*{-1em}
\end{figure}
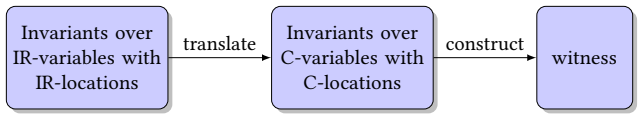

\noindent We cannot expect existing verification tools which we wish to use as helpers to be able to work on CFAs, to understand the safety property or to produce witnesses. Hence, we foresee two further sorts of components in our framework:
\begin{description}
  \item[Mapper]  
      A \textit{mapper} transforms the safety property specification inside the program into the desired input format of the helper. 
      A mapper basically conducts some simple syntactic code replacements.
     For instance, for our running example some helpers might instead require the safety property to be written as \texttt{ if(!(n==y))}\texttt{\{verifier\_error();\} } or  \texttt{ assert(n==y); }

  \item[Adapter] 
     An \textit{adapter} generates a correctness witness out of the computed loop invariants of a helper. 
     Furthermore, some \IG s work on intermediate representations (IR) of the C-language (e.g.~LLVM) 
     or intermediate verification languages (e.g.~Boogie).
     In this case, the computed invariants (formulated in terms of IR-variables) first of all need to be translated back to the namespace of the C-program. 
\end{description}

\noindent The latter transformations happening inside an adapter are shown in \Cref{fig:adapter}. 
Initially, the IR-language variables present in the invariants are translated to variables present in the C-program. 
After that, we transform their IR-code locations back to C-code locations.
For this, many compilers offer debug flags, adding this information to the IR.
Otherwise, building and matching the CFAs of the C-program and the IR-program is required.
Finally, the pairs of mapped location and invariant are stored in the form of a witness, constructable from the CFA.

\subsection{Cooperation within \covercig}
After having explained the individual components, we define their interaction in the framework.
In this paper, we focus on  the {\em parallel} execution of several helpers which implement complementary approaches so that 
we can leverage their individual strengths. 
Algorithm \ref{algorithm:cooperation} describes the form of cooperation. 
It is steered by several user configurable options which fix aspects like time and resource limits of master and helpers. 
\Cref{table:configurations} summarizes the configuration options. We next describe them in detail. 
\begin{table}
	\caption{Overview of the configuration options available} \label{table:configurations}
	\begin{tabulary}{\linewidth}{cCc}
		\toprule
	\textbf{Name} & \textbf{Description} & \textbf{Values}\\
	\midrule
	\texttt{restartMaster}		&restart the master after invariant generation & boolean\\
	\texttt{termAfterFirstInv}	&use first witness only & boolean\\
	\texttt{timerM}			&maximum time for master until \texttt{requestsForHelp} is send & time in s\\
	\texttt{timeoutH}			& maximum time for helpers to generate an invariant &time in s\\
	\bottomrule
	\end{tabulary}
\end{table}

\begin{description}
  \item[Master options]  The following aspects of the master's behavior need to be fixed: 
      First, when to delegate tasks to helpers, and 
      second, how to continue the verification process after invariant generation.
	  For the delegation, we let the \master \ run until it requests support, which  can be checked by inspecting the master's flag \texttt{requestsForHelp}.
	  The master gets a configurable timelimit (called \texttt{timerM}) after which it is expected to send this request. 
       By adding such an explicit request for help, we allow the master to send a request for other reasons (besides the timer) in the future. 
      Then, after invariant generation, the master can either be freshly restarted or continued (option 
      \texttt{restartMaster}).
  \item[Helper option] 
When at least two helpers run in parallel, eventually one of them first computes a witness. We can then either (1) directly stop the 
other helpers, or (2) wait for all to complete before injecting witnesses into the master. This option is called \texttt{termAfterFirstInv}. 
  
  \item[Timeouts] Finally, similar to the master, we can set a specific timeout for the helpers which fixes how long they
     are allowed to try to generate invariants. The timeout option is called \texttt{timeoutH}. 
 
\end{description}

\algrenewtext{For}[1] 
{\textbf{for each } #1   \textbf{do parallel}}
\begin{algorithm}[t] 
	\caption{CoVerCIG-algorithm\label{algorithm:cooperation}}
	\begin{algorithmic}[1]
		\Require C  \Comment{CFA}
             \Statex \hspace*{.4cm} P \Comment{safety property} 
		\Statex \hspace*{.4cm} M \Comment{master}
             \Statex \hspace*{.4cm} Helpers \Comment{set of helpers}
             \Statex \hspace*{.4cm} conf \Comment{configuration} 
		\Ensure $\omega$ \Comment{witness}
             \Statex \hspace*{.6cm} b \Comment{result} 
             \State M.start(C, P, conf.timerM); 
             \State wait until (M.requestsForHelp $\vee$ M.hasSolution());
             \If {(M.hasSolution())}
                 \State \Return M.getSolution();
             \EndIf
		\For{H $\in$ Helpers} \Comment{run helpers in parallel}  
                 \State H.start(C, P, conf.timeoutH);
                 \State wait until (H.timedout() $\vee$ H.hasSolution()  $\vee$ H.stopped()); 
                 \If{(H.hasSolution() $\wedge$ nonTrivial(H.getSolution()))}
                     \State witnesses := witnesses $\cup$ H.getSolution();
                     \If {(conf.termAfterFirstInv)} 
                           \For {H' $\in$ helpers $\setminus \{$ H $\}$} 
                              \State H'.stop(); \Comment{stop other helpers} 
                           \EndFor  
                     \EndIf
                 \EndIf
             \EndFor	
             \If {(M.hasSolution())}  
                \State \Return M.getSolution(); 
             \EndIf 
             \If {(witnesses $\neq \emptyset$)} \Comment{invariants found}  
                 \If {(conf.restartMaster)}
                      \State M.stop();
                  \EndIf
                  \State M.inject(witnesses); \Comment{inject witnesses into master}
                  \If{(conf.restartMaster)}
                     \State M.start(C,P, $\infty$);
                  \EndIf
             \EndIf 
             \State join(M); \Comment{wait for M to finish} 
             \State \Return M.getSolution();      
	\end{algorithmic}
\end{algorithm}

\noindent Next, we explain the CoVerCIG algorithm shown in Algorithm~\ref{algorithm:cooperation} in detail.
We assume that master and helpers run as threads and can be started and stopped.
We furthermore employ methods \texttt{wait} for waiting until some condition is achieved and 
\texttt{join} for waiting for a specific thread to complete. 

Initially, the \master \ is started without any \IG s running in parallel (line 1), providing the opportunity to verify programs on its own.
It runs standalone until it requests for help (either due to not being able to solve the problem alone or due to hitting its timer) or it computes a result which is subsequently returned (line 3). 
%
Afterwards all helpers are started in parallel (lines 5 and 6). They also run until they reach their timeout, a solution is found or they are stopped. 
Their solutions (invariants) are inserted into the witness set (line 9). Depending on option \texttt{termAfterFirstInv}, 
either all but the first finished helper are stopped or it is waited until all helpers either computed a solution or ran into their timeout. 
If invariants (witnesses) have been computed, these are injected into the master (line 18).  
If the \texttt{restartMaster} option is set, the master needs to be stopped before injection and restarted afterwards. 
Then the master continues and completes its verification (without any further request for help) and the result is finally returned.
%
\begin{example} 
	To explain the framework's functionality, we de\-monstrate the CoVerCIG algorithm on the example presented in \Cref{fig:example-a}.
	Assume that we instantiate the framework with a \master \ and four \IG s\footnote{Later, we will see that more than two helpers does not practically make sense.}.
	Moreover, we configure the framework as follows:
	\begin{itemize}
	\item restartMaster = true, 
	\item terminateAfterFirstInv = false,
	\item timerM = 50s, timeoutH = 300s.
	\end{itemize}
	Initially, the \master \ runs standalone and after 50 seconds runtime it requests help.
	This means that it cannot generate the invariant $n=x+y$, for example because it rather proves single program traces safe. 
	The \master \ would then run in parallel with the four \IG s being called. 
	Let us assume that the first helper returns only trivial invariants (after 10s), the second one an invariant $n\geq y$ (after 50s), the third one the invariant $n=x+y$ (after 100s) and the fourth the invariant $n-x-y=0$ (after 500s).
	The trivial invariant is ignored (see check in line 8) and when the second helper returns a solution, the third and fourth helper are still not stopped, due to the chosen configuration.
	The algorithm waits until the third helper computes the invariant and the fourth (only being able to compute an invariant after 500s) hits the timeout.
	Then the master is stopped, the invariants $n\geq y$ and $n=x+y$ are injected and the master is restarted.
	The \master\ can use both invariants and might now compute the correct result.
	%
	%
	%
	%
	%
\end{example}

\section{Implementation}

\begin{table*}[t]\caption{Summary of tools used as helpers} \label{table:tools}
\begin{tabular}{c c cc}
	\toprule
	\textbf{Tool} & \textbf{Techniques} & \textbf{Mapper} & \textbf{Adapter}\\
	\midrule
	 SeaHorn 				&	generation and solving of constrained horn clauses				&\ding{56}		&\ding{52}\\
	 UltimateAutomizer 				&	predicate abstraction, automata, path-based refinement				&\ding{56}		&(\ding{52})\\
	 VeriAbs		&	portfolio of 4 different sequential compositions 				&\ding{56}		&\ding{56}	\\
	\bottomrule	
\end{tabular}

\end{table*}

To be able to evaluate the performance of our framework \covercig, we instantiated it with two different \master s and 
three helpers, using existing off-the-shelf invariant generation tools. 
As verifiers need to be extended with witness injection for being able to act as a master, we used  the open-source configurable program analysis framework \cpachecker~\cite{DBLP:conf/cav/BeyerK11} for this purpose and employed two of its standard instantiations (predicate abstraction and k-induction). 
 We also decided to implement Algorithm~\ref{algorithm:cooperation} within 
\cpachecker. 
For the \IG s -- which can be used off-the-shelf -- we looked at current and past participants of the annual competition of 
software verification \svcomp~\cite{DBLP:conf/tacas/Beyer17}. Our intention was to find tools which provide complementary techniques 
for  invariant generation. To this end, we chose the tools \seahorn~\cite{DBLP:conf/cav/GurfinkelKKN15}, \ua~\cite{DBLP:conf/tacas/HeizmannCDGHLNM18} and \veriabs~\cite{DBLP:conf/kbse/AfzalACCDDKV19}. 
All \IG s are used as black-boxes.
An overview of the techniques employed in these tools is given in \Cref{table:tools}. The table also states whether the helpers require 
mappers and adapters. A more detailed explanation is given next.

\subsection{Master Verifiers}
\textbf{Predicate Abstraction.} 
The first analysis used as master is a predicate abstraction technique~\cite{DBLP:conf/fmcad/BeyerKW10}, conducting predicate refinement using a CEGAR (counter example guided abstraction refinement) scheme~\cite{DBLP:journals/jacm/ClarkeGJLV03} with lazy-abstraction~\cite{DBLP:conf/popl/HenzingerJMS02} and Craig interpolation~\cite{DBLP:conf/popl/HenzingerJMM04}.
Loop heads and error locations are used as locations where abstractions are computed; the computation of the abstraction itself  is done using an SMT solver. \\
\textit{Witness Injection:} For using this technique as master, we extended it with witness injection.
The purpose of witness injection is the use of the invariants as given in the witnesses in the {\em running} analysis 
of the master. 
It is realized by extracting predicates from the invariants and inserting them into the set of available predicates 
as maintained by the analysis. 
If these predicates contain conjunctions of clauses, these are furthermore split up and inserted individually.
Splitting predicates increases the performance due to the fact that SMT solvers perform better on many small predicates than on few larger ones\footnote{This has been reported by tool developers and has also shown in our experiments.}. 

\textbf{k-Induction}.
The basic idea of k-induction~\cite{DBLP:conf/sas/DonaldsonHKR11} is to generalize bounded model checking (BMC)~\cite{DBLP:journals/ac/BiereCCSZ03} via induction.
After proving k-bounded program executions safe using BMC, a generalization is aimed for.
The applied technique therefore  generates auxiliary invariants that are continuously refined using a CEGAR based analysis~\cite{DBLP:conf/cav/0001DW15}.	
These invariants are combined with the information generated by BMC and generalized to a safety proof by successfully conducting an induction step.
\\ 
\textit{Witness Injection:}
We also implemented a method for witness injection for k-induction.
Witness injection before restarting the analysis is conducted by proving correctness of witnesses and afterwards adding them to the set of invariants maintained by the analysis.
For injecting witness into a running analysis, we periodically check   whether new witnesses are available before each induction step.
If so, we also check their validity and add them to the set of invariants.

\subsection{\InvGen s}
We have chosen the existing verification tools \seahorn, \ua \ and \veriabs \ as \IG s. 
For this choice, we have inspected the tools participating in the annual competition on software verification. 
\ua \  has achieved excellent results in this year's \svcomp \ by taking the second place in the category ''Overall''.
Since we already use two main components of \cpachecker, the winner in the category ''Overall'' this year, inside the masters, we did not also take 
instantiations of \cpachecker \ as helpers into account. 
Moreover, we have chosen \veriabs, the winning tool in \svcomp \ 2020  on ''ReachSafety'', the category dealing with safety properties.
As third tool we use \seahorn , a verification tool based on constrained Horn Clauses, neither currently participating  in the \svcomp \ nor producing witnesses.
The three \IG s employ verification techniques complementary to those of both the other helpers and the two masters. 
%

\textbf{\seahorn.} 
\seahorn~\cite{DBLP:conf/cav/GurfinkelKKN15} is a verification tool using Constrained Horn Clauses (CHCs) to solve the verification tasks.
\seahorn \ constructs CHCs for each statement, encoding both data and control dependencies as well as the safety property.
The (recursive) system of CHC is solved using the solver \textsc{Spacer}~\cite{DBLP:conf/cav/KomuravelliGC14}.
\textsc{Spacer} tries to prove the unsatisfiability of the CHCs, being equivalent to proving the program safe, by searching for interpretations of the predicates present in the CHCs.
\seahorn \ operates on the LLVM intermediate representation.
We choose \seahorn \ to extend the stack of helpers  by a tool being conceptually complementary to the others. 

\textit{Adapter for \seahorn.} \seahorn \ participated in the \svcomp \ 2015, thus it can process the encoding of safety properties used in our evaluation.
Unfortunately, it only returns a boolean answer and no verification witnesses, hence we had to implement an adapter for it. 
Our adapter follows the general construction explained in \Cref{fig:adapter} and we exemplify its translation in \Cref{example:BasicBlock}.
\begin{example}
	\label{example:BasicBlock} 
	SeaHorn associates invariants to LLVM basic blocks. 
	A basic block\footnote{https://releases.llvm.org/5.0.0/docs/LangRef.html\#functions} is a code fragment having a single entry location (the first) and a single exit location (in general the last location of the block).  
	We obtain the computed invariants in LLVM and the corresponding basic blocks by using the launch parameter \texttt{$-$$-$show-invars}.
	To construct a witness containing them, we need to translate the invariants and find the matching C-code location for the basic block.
	For both, we use the LLVM-IR equipped with debug information, using \seahorn   with launch parameter \texttt{-g} .
	Thereby, we obtain the IR-code fragment of the program in \Cref{fig:example-a}, shown in simplified form and containing the most important debug information as comments.
	The example contains two Basic Blocks, \texttt{entry} and \texttt{\_bb}.	
	\begin{lstlisting}[tabsize=2, columns=flexible,mathescape=true,numbers=left,basicstyle=\ttfamily,	stepnumber=1, language=LLVM,xleftmargin=.1\linewidth]
entry:
v1 = bitcast i32 (...)* @nondet to i32 ()*   $\ \ \  \triangleright n$
v2 = icmp eq i32 v1, 0
br i1 v2, label %error, label %_bb

_bb:                                          
v3 = phi i32 [0, %entry], [v6, %_bb]     $\ \ \ \ \ \ \  \ \  \triangleright y$
v4 = phi i32 [v1, %entry], [v5, %_bb]  $ \ \ \ \ \ \  \ \  \triangleright x$
v5 = add i32 v4, -1
v6 = add i32 v3, 1
v7 = icmp eq i32 v5, 0
br i1 v7, label %error, label %_bb $\ \  \ \ \ \ \  \triangleright \mathit{line~ 4} $
$\cdots$
	\end{lstlisting}
	\seahorn \ computes the invariant $v1-v4-v3 = 0$ for the example and associates it with the basic block \texttt{\_bb}.
	At first, we need to transform the variables from the IR to C-variables occurring in the program.
	In this example we can use the debug information, as shown in comments in the code.
	In general, a more sophisticated procedure is needed since LLVM-IR uses a three address code.
	Therein, complex expressions are split into several statements using intermediate variables which are resolved to C-expressions.
	
	Afterwards, the transformed invariant needs to be associated with the correct location in the C-code. 
	We analyze the LLVM IR program structure to map the basic blocks back to C-locations.
	In the example, the block \texttt{\_bb} is identified as being the loop of the program, thus the invariant is mapped to the loop head.
	For this, we employed some basic functions provided by \textsc{PHASAR}~\cite{DBLP:conf/tacas/SchubertHB19} in our adapter.
	Finally, we construct the CFA of the C-program, store the invariants at the nodes and convert the equipped CFA to a verification witness.
	
\end{example}

\textbf{\ua.} 
\ua's verification technique is based on predicate abstraction and on automata constructions~\cite{DBLP:conf/cav/HeizmannHP13,DBLP:conf/tacas/HeizmannCDGHLNM18, DBLP:conf/tacas/HeizmannCDGNMSS17}.
The program is represented as finite automaton and error labels are final states.
\ua \ then aims at proving emptiness of the accepted language of the automaton which is equivalent to proving safety of the  program.
Although \ua \ produces verification witnesses, we added an adapter for the witnesses due to currently existing technical incompatibilities.

\textbf{\veriabs.} 
\veriabs \ is using a portfolio of four different verification techniques, each containing several sequentially composed components~\cite{DBLP:conf/kbse/AfzalACCDDKV19}.
The selection of strategies (techniques) from the portfolio is performed by analyzing the loop structure and intervals for variables used in the loop.
Depending on the analysis result, one of the following four techniques is applied:
(1) random fuzz testing, (2) techniques to abstract arrays and apply BMC afterwards, (3) explicit state model checking followed by standalone invariant generation techniques or (4) a fixed sequence of different verification approaches. 
%

\section{Evaluation}
In the following, we evaluate different instantiations of \covercig.
We focus on both effectiveness and efficiency, generally aiming at checking whether the use of \covercig \ can increase the number of correctly solved verification tasks within the same resource limits.

\subsection{Research Questions}

We start with the feasibility of the approach in general. 

\textbf{Feasibility hypothesis:}
A framework for collective invariant generation can be constructed using existing tools by building adapters when needed.
\textit{Evaluation plan:}
We construct instances of the framework, using instances of predicate abstraction and k-induction as \master \ and using three off-the-shelf \IG.
As a result, we obtain 14 different combinations.

%
Besides feasibility, we were interested in the following four research questions. 
\begin{description}
\item[RQ1.]
Can collective invariant generation increase the effectiveness of the master verifier? 
\textit{Evaluation plan:} We let the framework run with a single invariant generator and compare the results to a run where the \master \ runs standalone.

\item[RQ2.]
Does cooperation impact the overall efficiency of the  verification? 
\textit{Evaluation plan:} We compare the run time of \covercig \ with one helper against the two \master s running standalone.

\item[RQ3.]
What is an appropriate time for the master to run before requesting for help?
\textit{Evaluation plan:} We run \covercig \ in combination with one helper, evaluating the effectiveness of requesting for help after 50, 100 and 200 seconds.

\item[RQ4.]
Does it pay off to run two invariant generators in parallel? 
\textit{Evaluation plan:} We let the framework run with two invariant generators and compare the results to a run, where only a single invariant generator is used.
Moreover, we compare the two configurations for \texttt{termAfterFirstInv} and evaluate timeouts for helpers using 100s and 200s.

\end{description} 

\subsection{Experimental Setup}
\textbf{Tools.}
We based the implemented of our \covercig \ algorithm on the \cpachecker\footnote{https://github.com/sosy-lab/cpachecker} 1.9.1 (8646a85) using  MathSat5\footnote{https://mathsat.fbk.eu/} as solver within \cpachecker.
For the helper \veriabs \ and \ua \ we used the versions as used in the \svcomp \ 2020\footnote{https://gitlab.com/sosy-lab/sv-comp/archives-2020/tree/master/2020}.
Due to the fact that there is no precompiled binary of \seahorn, we employ the docker container of the latest version\footnote{suggested by the developers; used docker seahorn/seahorn-llvm5 (4c01c1d)}.
All three \IG s are used in their default configuration.

During evaluation, we used the following default configurations for our framework:
We set \texttt{termAfterFirstInv} and \texttt{restartMaster} to true, setting the \texttt{timerM} to 50s and the \texttt{timeoutH} to 300s.
The master and helper used in a specific configuration as well as changes made to the default configurations are denoted as follows:
The configuration \texttt{kInd-ua-va-100}\texttt{-wait-200} denotes a configuration using k-induction as master and the helpers \ua \ and \veriabs.
The \texttt{timerM} is set to 100s, \texttt{termAfterFirstInv} to false and \texttt{timeoutH} to 200s.
In general, we will use the abbreviations SH for \seahorn, UA for \ua \ and VA for \veriabs.

\textbf{Verification Tasks.}
The verification tasks used are taken from the set of \svcomp \ 2020 benchmarks\footnote{https://github.com/sosy-lab/sv-benchmarks/releases/tag/svcomp20}.
As we are interested in finding suitable loop invariants, we selected all tasks from the category ReachSafety-Loops.
To obtain a more broad distribution of tasks, we randomly selected 55 additional tasks from the categories ProductLines, Recursive, Sequentialized, ECA, Floats and Heap, yielding in total 342 tasks.

\textbf{Computing Resources.}
We conducted the evaluation on three virtual machines, each having an Intel Xeon E5-2695 v4 CPU with eight cores and a frequency of 2.10 GHz and 16GB memory, running an Ubuntu 18.04 LTS with Linux Kernel 4.15.
We run our experiments using the same setting as in the \svcomp, giving each task 15 minutes of CPU-time on 8 cores and 15GB or memory.
We employed \benchexec \ thereby guaranteeing the resource-limitations~\cite{DBLP:journals/sttt/BeyerLW19}.
All experimental data are available\footnote{https://covercig.github.io/}.

\subsection{Experimental Results}
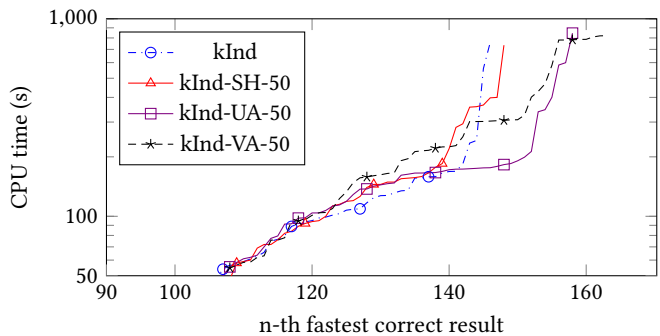
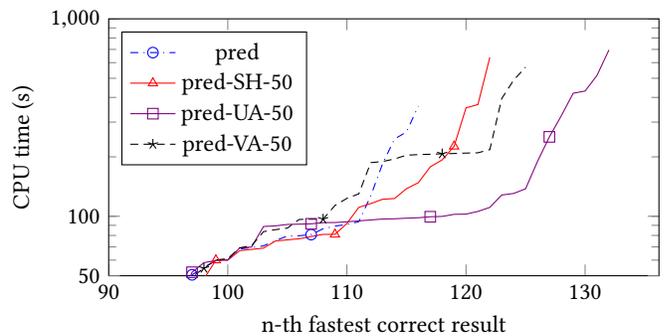
\begin{figure*}[th]
	\begin{subfigure}{.5\textwidth}
		
		\begin{tikzpicture}
		\begin{semilogyaxis}[
		height=5cm, width = \linewidth,
		/pgfplots/table/y index=3,
		/pgfplots/table/header=false,
		xlabel=n-th fastest correct result,
		ylabel=CPU time (s),
		xmin=90,
		ymin=50,
		ymax=1000,
		mark repeat=10,
		mark options={solid},
		extra y ticks={50},
		extra y tick labels={50},
		extra x ticks={90},
		extra x tick labels={90},
		legend entries={kInd, kInd-SH-50, kInd-UA-50, kInd-VA-50},
		every axis legend/.append style={at={(0,1)}, anchor=north west, outer xsep=5pt, outer ysep=5pt, }
		]
		\addplot+[dashdotted,mark=o] 						table {figures/rq1/kInd.csv};
		\addplot+[ mark=triangle,	] 	table {figures/rq1/kInd-sh-50.csv};
		\addplot+[mark=square, violet] 					table {figures/rq1/kInd-ua-50.csv};
		\addplot+[densely dashed,black	] 						table {figures/rq1/kInd-va-50.csv};
		\end{semilogyaxis}
		\end{tikzpicture} 
		\caption{\covercig \ using k-induction as master} \label{fig:quantile-rq1-kInd}
		
	\end{subfigure}
	\nolinebreak
	\begin{subfigure}{.5 \textwidth}
		\begin{tikzpicture}
		\begin{semilogyaxis}[
		height=5cm, width = \linewidth,
		/pgfplots/table/y index=3,
		/pgfplots/table/header=false,
		xlabel=n-th fastest correct result,
		ylabel=CPU time (s),
		xmin=90,
		ymin=50,
		ymax=1000,
		mark repeat=10,
		mark options={solid},
		extra y ticks={50},
		extra y tick labels={50},
		legend entries={ pred, pred-SH-50, pred-UA-50, pred-VA-50},
		every axis legend/.append style={at={(0,1)}, anchor=north west, outer xsep=5pt, outer ysep=5pt,}
		]
		\addplot+[dashdotted,mark=o] 		table {figures/rq1/pred.csv};
		\addplot+[ mark=triangle,	] 	table {figures/rq1/pred-sh-50.csv};
		\addplot+[mark=square, violet] 	table {figures/rq1/pred-ua-50.csv};
		\addplot+[densely dashed,black	] 		table {figures/rq1/pred-va-50.csv};
		\end{semilogyaxis}
		\end{tikzpicture}
		\caption{\covercig \ using predicate abstraction as master}
	\label{fig:quantile-rq1-pred}
	\end{subfigure}
		\caption{Quantile plots for \covercig \ using both masters and different single helpers.} \label{fig:quantile}
\end{figure*}
\textbf{Feasibility hypothesis.}
We implemented the \covercig -frame\-work as proof-of-concept in the \cpachecker-framework.
For this, we had to extend the existing implementations of k-induction and predicate abstraction with witness injection.
For the \IG s we did not change a single line of code, only adding adapters for \seahorn \ and \ua.
Integrating helpers like \veriabs , not requiring an adapter or a mapper, can be done within a few lines of code.
Although the implementation is a proof-of-concept, this shows that the presented framework works in practice and is applicable to all kinds of off-the-shelf \IG s, those producing verification witnesses and those generating invariants in IR.

\textbf{RQ1 (Effectiveness).}
To evaluate whether a \master \ benefits from the support of a helper, we execute a combination of a master and a helper in the default configuration and compare it to the master running standalone. Here, we are interested in the number of {\em correct} verification results, 
i.e., the verifier correctly reporting the safety property to be fulfilled (result $true$) or not (result $\mathit{false}$). 
Running standalone, k-induction can correctly solve 146 of the verification tasks, predicate abstraction 116.

\begin{table}[t]
	\caption{Comparison of the two \master s running standalone and using a single helper.}
	\begin{tabulary}{\linewidth}{c c C C C C  }
		\toprule
	\textbf{	Tool-}	&  \multicolumn{3}{c}{\textbf{correct} }   &\multicolumn{2}{c}{\textbf{additional} }  \\
		\textbf{Combination}&\textbf{overall} & \textbf{true}&\textbf{false}&\textbf{true}&\textbf{false}\\
		\midrule
		k-induction			&146		&102	& 44 	&-		&-		  \\
		kInd-SH-50			&148		&104	& 44 	&+3		&0		\\	
		kInd-UA-50			&158		&114	& 44	&+13	&0		\\	
		kInd-VA-50			&163		&119	& 44	&+19	&0		\\	
		\midrule
		pred abstr.			&116		&78		& 38	&-		&-	  \\
		pred-SH-50			&122		&84		& 38	&+6		&0	  \\
		pred-UA-50			&132		&94		& 38	&+16	&0	  \\
		pred-VA-50			&125		&87		& 38	&+9		&0	\\	
		\bottomrule
	\end{tabulary}
	\label{tab:results-rq1}
\end{table}

Table~\ref{tab:results-rq1} gives the results of this experiment. In the table we see the overall number of correct results, 
the number of correct $true$ and correct $\mathit{false}$ results plus the the number of tasks additionally solved when using a helper.
Through the cooperative invariant generation, the performance of both masters is increased. 
As expected, this applies to verification tasks with fulfilled safety property only, i.e., the invariant generators 
can help in proving a property to hold, but cannot help in refuting properties (as they correctly do not generate invariants in these cases). 
Besides the additionally solved tasks, there is also one (for SH and UA) and two (for VA) tasks, respectively, which cannot be correctly solved anymore. 
In these cases, the master alone consumes nearly all of the CPU time available, hence sharing resources in cooperation  with the helpers results in a timeout.

%

\llbox{On our data set, the total number of correctly solved tasks increases using \covercig \ by 12\% for k-induction and 14\% for predicate abstraction used as master.}

\textbf{RQ2 (Efficiency).}
Next, we evaluate the efficiency of \covercig, analyzing the CPU-time spend solving the verification tasks.
As \covercig \ eventually shares the CPU time between master and helpers, we expect that more time is needed to compute a correct result after the helper is started.

\Cref{fig:quantile} shows two quantile plots of the verification runs, the left with k-induction and the right with predicate abstraction as master. 
A datapoint $(x,y)$ in the plot means that the verifier computes the x-fastest correct results (for a task) in at maximal $y$ seconds.
As \covercig \ instances behave like masters standalone in the first 50 seconds, we only show results {\em not} solved within these 50 seconds. We see that for tasks requiring a low amount of time, all instances (including the master alone) require a similar amount of CPU time. 
For tasks requiring more time, \covercig \ is actually often faster, the extreme being predicate abstraction as master which alone is unable 
to solve more difficult tasks in the given time. 

\begin{figure}
	\scalebox{1}{
	\begin{tikzpicture}
	\begin{loglogaxis}[
	xlabel=k-induction standalone (s),
	ylabel=\covercig \ with kInd-\veriabs-50 (s),
	xmin=1,
	xmax=1000,
	ymin=1,
	ymax=1000,
	domain=1:1001,
	clip mode=individual,
	axis equal image,
	]
	\addplot+[blue, mark=+,only marks]
	table[
	header=false,
	skip first n=3, 
	x index=2, 
	y index=6  
	] {figures/scatter-va.table.csv};
	\addplot[gray] {x};
	\addplot[red, dashed] coordinates {(50,1) (50,50) (1,50)};
	\end{loglogaxis}
	\end{tikzpicture}}
	\caption{Scatter plot comparing kInd and kInd-\veriabs-50} \label{fig:scatter-kInd}
\end{figure}
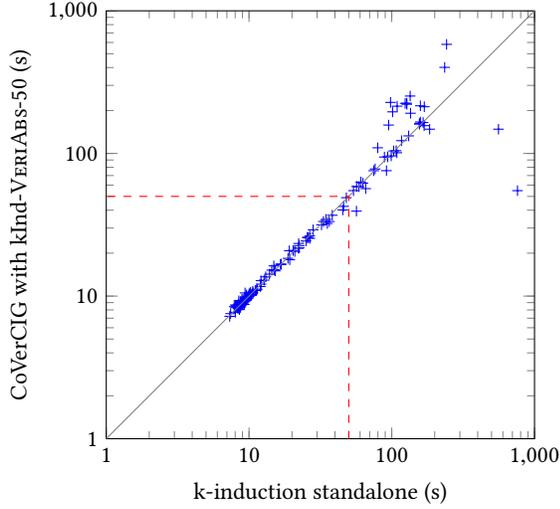

We exemplarily also compared the CPU time of k-induction standalone with \covercig \ using \veriabs \ as helper {\em per task}. 
It turns out that sharing does only slightly impact the runtime, as shown in \Cref{fig:scatter-kInd}.
The scatter plot compares the CPU time of k-induction standalone as master and k-induction supported by \veriabs, in case both tools solved the task correctly.
A datapoint $(x,y)$ means that k-induction standalone takes $x$ seconds to solve the task and in combination with \veriabs \ $y$ seconds.
The red dashed box contains all tasks solved within 50 seconds, where both tools behave equally, since the master does not request for help in these cases. We see some tasks for which helping increased the runtime, but also some for which it decreased it. 
In most of the cases, the CPU time used by \covercig \ is not significantly higher.

Finally,  we compare the average CPU time needed to correctly solve a task.
Table~\ref{tab:runtime} shows the average time needed for all tasks and -- in brackets -- for the correctly solved tasks only. 
We observe that the runtime increases when only looking at correctly solved tasks (in particular for \veriabs), however,
 when considering all tasks the CPU time is even decreased. The latter effect is due to the number of timeouts of the master 
decreasing when cooperating with helpers. 
Concluding, we can make the following observation. 
\begin{table}[t]
	\caption{ Total CPU time for all tasks and average CPU time taken for a correct answer in brackets, both in seconds.}	\label{tab:runtime} 
	\begin{tabular}{l cc cc}
		\toprule
		&	\multicolumn{1}{c}{\textbf{Master}}		& \multicolumn{3}{c}{\textbf{Master} }\\
		&	\multicolumn{1}{c}{\textbf{standalone}}  & \multicolumn{1}{c}{\textbf{+SH}\tablefootnote{Due to possible imprecisely measured CPU time of \benchexec, we computed an upper bound on the runtime.}} &\multicolumn{1}{c}{\textbf{+UA}} 	&\multicolumn{1}{c}{\textbf{+VA}}\\
		\midrule		
		kInd				&	491 	(50)					&489  (63)				&477  (68)				&482  (107)	 \\
		Pred				&	479 	(30)					&468  (39)					&454  (51)				&470  (49)\\
		\bottomrule
	\end{tabular}
\end{table}


\llbox{On our dataset, collaborative invariant generation does not negatively impact the effectiveness; 
in some cases we even see small improvements. }

\textbf{RQ3 (Time for the Master to request for help).} 
To determine a preferable time for the master to run alone, we evaluated \covercig \ using 50, 100 or 200 seconds for \texttt{timerM}.
A summary of the results is given in \Cref{tab:results-timer}, showing the number of correctly solved tasks for each instantiation.
\begin{table}
	\caption{Number of correctly verified tasks for different parameters of \texttt{timerM}} \label{tab:results-timer}
	\begin{tabulary}{\linewidth}{C CCC  CCC}
	\toprule
	\textbf{Value}  &\multicolumn{3}{c}{\textbf{k-induction}}&\multicolumn{3}{c}{\textbf{predicate abstr.}}\\
	\texttt{\textbf{timerM}}	&\textbf{-SH} 	&\textbf{-UA} 	&\textbf{-VA}	&\textbf{-SH} 	&\textbf{-UA} 	& \textbf{-VA}	\\
	\midrule	
	50s				&148	&158	&163	&122	&132	&125\\
	100s			&148	&157	&161	&122	&132	&125\\
	200s			&148	&157	&156	&122	&132	&125\\
	\bottomrule
\end{tabulary}

 \end{table}
Both masters achieve their best result running alone for 50 seconds. 
For k-induction, a good choice for \texttt{timerM} plays an important role for its performance.
In contrast, the  results of predicate abstraction are not influenced by different values for \texttt{timerM} at all, because predicate abstraction computes its correctly given answers on average in 19 seconds after obtaining the invariants by the helpers.
When using k-induction, we observe cases where the correct solution is computed only if the master sends the request early.
Asking later sometimes leads to a situation where the invariant is obtained too late to be helpful. 
Hence, we employed 50 seconds in our default configuration which we used to evaluate RQ1 and RQ2.  
\llbox{On our dataset, \covercig \ performs best when requesting early for help, using 50 seconds for \texttt{timerM}.} 


\smallskip 
\textbf{RQ4 (Combination of helpers).}
In RQ4, we were interested in finding out (a) whether it is beneficial to run two invariant generators in parallel, and (b) 
if yes, which pair is best for this.  

\begin{figure}[t]
	\begin{subfigure}{.5\linewidth}
	\includegraphics[scale=0.26	]{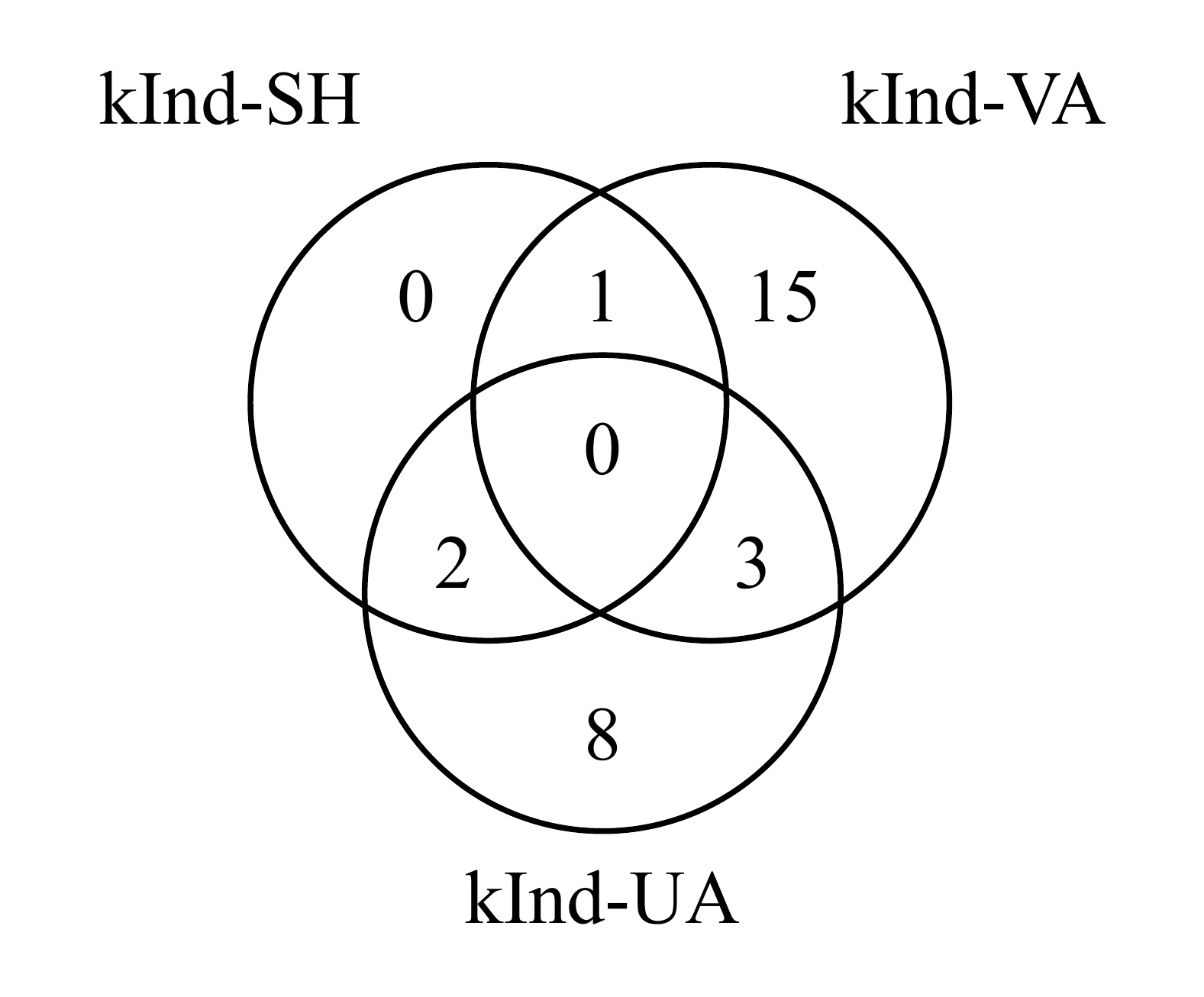}
		\caption{For k-induction}
	\end{subfigure} \nolinebreak
	\begin{subfigure}{.5\linewidth}
	\includegraphics[scale=0.26	]{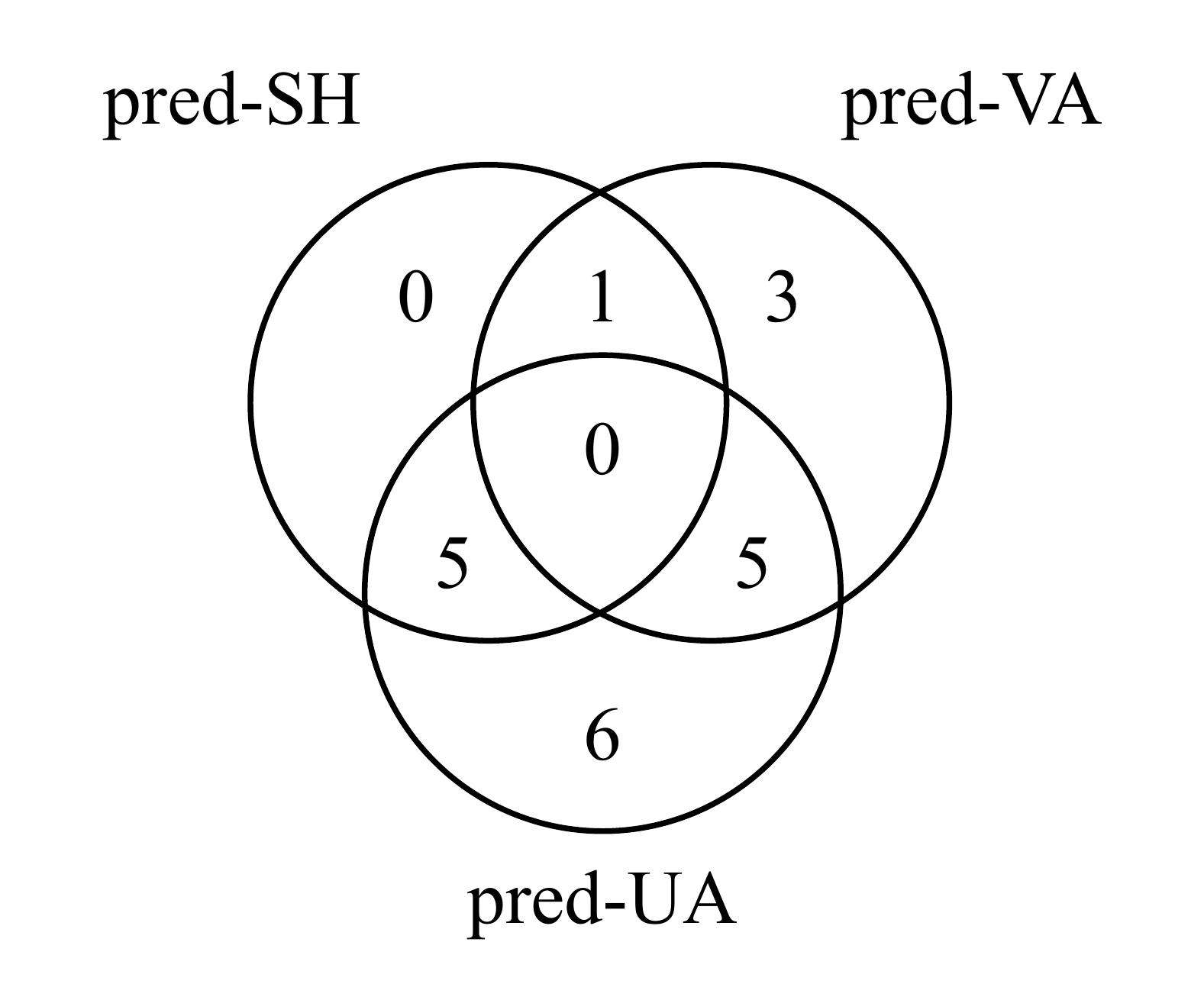}
		\caption{For predicate abstraction}
	\end{subfigure}
	\caption{Tasks additionally solved using single helpers}\label{fig:venn} 

\end{figure}

To this end, we first of all determined  which helper is able to solve which of the additionally solved tasks. 
The result is shown in the Venn diagrams of \Cref{fig:venn}. 
Surprisingly, \seahorn \ -- although employing a technique conceptually different to \ua \ and \veriabs \ -- is not able to solve a single task which not at least one of the others can. 

Next, we thus studied the number of correctly solved tasks using the three possible pairs of helpers,  
running the two helpers in a pair in parallel. \Cref{table:cooperation} in the first row shows the results. 
It in addition also contains results evaluating two values (100 and 200 seconds) for parameter \texttt{timeoutH} 
in a setting when the master waits for all helpers to complete (not just the first one). 
A first observation is that -- except for the case of k-induction with UA-VA -- the results show no significant difference when 
using the default configuration or wait-100 or wait-200.

For checking whether parallel execution of helpers is beneficial, these numbers need to be compared against those for 
a single helper as given in Table~\ref{tab:results-rq1}. 
We see that predicate abstraction benefits from using two helpers, especially using \ua \ and \veriabs.
Using \covercig \ with these tools perfectly combines their strengths,  
thereby  increasing the number of correctly solved tasks in total by 17\%.
In contrast, it turns out that for k-induction none of the combinations of two helpers outperforms \covercig \ using \veriabs\ only.
For \ua \ and \veriabs\  as helpers, the total number does not change, only the set of solved tasks. 
For instance, nearly 50\% of the additional tasks solved by kind-UA-VA are not solved using kInd-UA and vice versa.
This result is based on the fact that they have to share the available CPU time in the combination.
Hence, tasks that are solved using one of them as helper alone could not be solved anymore in a combination because of timeouts. 
This phenomenon is even more an issue when running all three helpers in parallel.
The combination of all three helpers solves only 154 tasks correctly for k-induction and 129 for predicate abstraction.
\llbox{On our dataset, \covercig \ can increase the total number of correctly solved tasks using UA and VA in parallel;
in general waiting for the other tool to also finish its computation does not pay off.} 
\begin{table}[t]
	\caption{Number of correctly solved tasks using different forms of cooperation with two helpers running in parallel.}\label{table:cooperation}
	\begin{tabulary}{\linewidth}{c CCC  CCC}
		\toprule
		\textbf{Config} &\multicolumn{3}{c}{\textbf{k-induction}}&\multicolumn{3}{c}{\textbf{predicate abstr.}}\\
		&\textbf{SH-UA }	&\textbf{SH-VA} 	& \textbf{UA-VA}	&\textbf{SH-UA }	&\textbf{SH-VA 	}&\textbf{ UA-VA}	\\
		\midrule	
		default		&153	&156	&163	&130	&130	&136\\
		wait-100	&156	&155	&161	&132	&131	&136\\
		wait-200	&155	&156	&155	&132	&129	&135\\
		\bottomrule
	\end{tabulary}
\end{table}

\subsection{Threads to Validity} \label{sec:validity}
We have conducted our evaluation using a random sample of tasks as well as those in the category Loops. 
Although this guarantees some diversity in the chosen tasks, our findings may not completely carry over to arbitrary real-world programs.

The experiments are conducted using the reliable framework \benchexec\ on identical machines with same resource limitations, guaranteeing comparable results.
As \seahorn \ is used within a docker-container, its CPU usage however cannot be measured by \benchexec.
We therefore measured its CPU usage externally, rounded it up and added it to the measured CPU time, obtaining a lower bound for the correctly solved tasks.
Thereby, all results stay valid, especially of the best performing instantiations of \covercig, as they do not use \seahorn.

Our implementation of \covercig\ relies on the correctness of the used \master s and helpers  (which are given) as well as on the adapters (which we build).
An incorrectly translated invariant may however influence the performance only negatively. 

Both \master s used as well as \ua \ and \veriabs \ are participating in the annual \svcomp, hence they might be tuned to the tasks employed. This does however not influence the validity of the results since our interest is in the {\em additional} number of  
tasks solved by cooperation, not the solved ones per se.  
\section{Related work}
In this paper, we presented a framework for cooperative verification via collective invariant generation. 
The idea of collaboration for verification by combining known techniques has been widely employed before.
For instance, there are combinations of verification with testing approaches~\cite{DBLP:conf/icse/CsallnerS05,csallner08dsd-crasher,DBLP:conf/icse/GeTXT11,DBLP:conf/fm/ChristakisMW12,DBLP:conf/icse/Christakis0W16,DBLP:conf/vmcai/DacaGH16} 
and with approaches for invariant generation~\cite{DBLP:conf/vmcai/SankaranarayananSM05, DBLP:conf/cav/GuptaR09,DBLP:conf/tacas/RochaRIC017,DBLP:conf/pldi/BlanchetCCFMMMR03,DBLP:conf/sas/Brain0KS15}. 
The latter combinations are conducted in a {\em white box} manner using strong coupling between the components, making the addition of a new approach a challenging task.
Our framework conceptually decouples the invariant generation from the verification, making it more flexible.
In addition, using a black box integration with defined exchange formats allows us to easily exchange or integrate new approaches.

There are also existing concepts for collaboration between different techniques in a {\em black-box} manner.
Conditional model checking is a technique for sequentially composing different model checkers, sharing information between the tools in form of conditions~\cite{DBLP:conf/sigsoft/BeyerHKW12}.
Beyer and Jakobs developed a concept for combining model checking with testing~\cite{DBLP:conf/fase/BeyerJ19}.
Although both approaches enable cooperation, none combines a verification tool and tools for invariant generation.

We next shortly discuss three approaches which are conceptually closer to our framework.
Frama-C is a framework for code analysis, 
aiming for analyzing industrial size code~\cite{DBLP:journals/fac/KirchnerKPSY15}.
The framework contains different plugins, each implementing a verification or testing technique.
The plugins can exchange information in form of ASCL source code annotations.
Within Frama-C, the analyzers can collaborate by being either sequentially or parallelly composed.
For this, partial results produced by an analysis can be completed by a second one  or several partial results computed in parallel are composed to a complete result.
Frama-C offers the general possibility to define cooperation between existing plugins.
To the best of our knowledge, Frama-C does however not provide a conceptual collaboration of a verification approach and tools for invariant generation driven by the verification approach's demand for support. 

The approach of using continuously refined invariants for k-induction~\cite{DBLP:conf/cav/0001DW15} uses a lightweight dataflow analysis 
which can be considered to be a helper for verification.
Therein, the supporting invariant generator runs in parallel to the k-induction analysis.
Compared to our framework, the main difference is the form of cooperation used. 
Beyer et al.~use a white-box integration for the cooperation between k-induction and the invariant generator, building hardly wired connections between both analyses and sharing the information {\em inside} the tool.
Thus, integrating external tools is hard to achieve. 
Moreover, the approach is designed to work for k-induction only.
Note that an analogeous approach is proposed by Brain et al.~\cite{DBLP:conf/sas/Brain0KS15}.

Pauck and Wehrheim proposed \textsc{CoDiDroid}, a framework for cooperative taint flow analysis for \android \ apps~\cite{DBLP:conf/sigsoft/PauckW19}.
Within their framework, different analysis tools with specialized capabilities are combined as black-boxes.
\textsc{CoDiDroid} is however tailored to the needs of \android \ taint flow analysis, thus the exchanged information differs.
Thus \textsc{CoDiDroid} is not able to orchestrate or exchange information on safety analysis with shared invariant generation.

To summarize, there are a lot of existing approaches for cooperative verification, but most of them are white-box combinations, 
and the only existing black-box combinations are not general enough to allows for collective invariant generation.

%
%
%
%
%

%
%

	\section{Conclusion}
In this paper, we have presented a novel form of black box cooperation for software verification via collective invariant generation.
Within the configurable framework named \covercig, the so called  \master \  steering the verification process is able to delegate the task of invariant generation to one or several \IG s.

We implemented \covercig \ within the \cpachecker\ framework using k-induction and predicate abstraction as master analysis supported by three existing helpers \seahorn, \ua \ and \veriabs.
Our evaluation on a set of  \svcomp \ verification tasks shows that \covercig \ increases the number of correctly solved tasks without increasing the overall verification time.
The best combination of helpers, \ua \ and \veriabs\ in parallel, yields an increase of 12\% for k-induction and 17\%  for predicate abstraction.

Next, we plan to enhance the cooperation by analyzing the behavior of the master in order to identify an optimal point to request for help.
Moreover, extending \covercig \ by additionally taking error traces found by the helper into account is also scheduled.
In addition, we intend to investigate whether a selection of helpers on the basis of the given verification task is beneficial. 


\bibliographystyle{ACM-Reference-Format}
\bibliography{literature}

\end{document}